\documentclass[prd,amsmath,amssymb,aps,showpacs,nofootinbib,10pt]{revtex4}

\usepackage{color,graphicx}
\usepackage{epstopdf}

\begin{document}

\title{
Production of the bottom analogues and the spin partner of the
{\boldmath$X(3872)$}\\ at hadron colliders
}

 \author{Feng-Kun Guo$^a$ }
 \author{Ulf-G. Mei{\ss}ner$^{a,b}$ }
 \author{Wei Wang$^{a,c}$}
 \author{Zhi Yang$^a$}

\affiliation{
  $^a$ Helmholtz-Institut f\"ur Strahlen- und Kernphysik and Bethe Center for
Theoretical Physics,
Universit\"at Bonn, D-53115 Bonn, Germany\\
$^b$ Institute for Advanced Simulation, Institut f\"ur Kernphysik and J\"ulich
Center for Hadron Physics,
JARA-FAME and JARA-HPC, Forschungszentrum J\"ulich, D-52425 J\"ulich, Germany\\
$^c$INPAC, Department of Physics and Astronomy, Shanghai Jiao-Tong University, Shanghai, 200240, P. R. China}

\begin{abstract}
Using the Monte Carlo event generator tools  Pythia and Herwig, we
simulate the production of  bottom/charm  meson and antimeson pairs  at
hadron colliders in proton-proton/antiproton collisions. With these results,
we derive an order-of-magnitude   estimate for the production rates of  the
bottom analogues and the spin
partner of the $X(3872)$ as  hadronic molecules  at the LHC
and Tevatron experiments. We find that  the cross sections for these
processes  are at the nb level, so that the current and future data sets from
the Tevatron and LHC experiments offer a significant discovery potential. We
further point out that the $X_b/X_{b2}$ should be reconstructed in the
$\gamma \Upsilon(nS) (n=1,2,3)$, $\Upsilon(1S)\pi^+\pi^-\pi^0$,  or
$\chi_{bJ}\pi^+\pi^-$  instead of  the $\Upsilon(nS)\pi^+\pi^-$  final states.

\end{abstract}

\maketitle

\section{Introduction}

As the $B$ factories and high energy hadron colliders  have accumulated
unprecedented data samples, a dramatic progress has been made in hadron
spectroscopy in the past decade. Especially, in the mass region of heavy
quarkonia, a number of new and unexpected structures have been discovered  at these
experimental facilities. Many of them defy an ordinary charmonium
interpretation,  among which the $X(3872)$ has received the most intensive
attention~\cite{Brambilla:2010cs} so far.

The $X(3872)$ was first discovered  by the Belle Collaboration in $B$ decays at
the $e^+e^-$ collider located at KEK~\cite{Choi:2003ue} and later confirmed
by the BaBar Collaboration~\cite{Aubert:2004ns} in the same channel.
It can also be copiously produced in high energy proton-proton/antiproton
collisions at the Tevatron~\cite{Abazov:2004kp,Aaltonen:2009vj} and
LHC~\cite{Chatrchyan:2013cld,Aaij:2013zoa}.  This meson is peculiar in several
aspects, and its nature is still under debate. The total width  is  tiny
compared to typical hadronic widths and only an upper bound has been set:
$\Gamma<1.2$~MeV~\cite{Beringer:1900zz}. The mass lies in the extreme close
vicinity to the  $D^0\bar D^{*0}$ threshold,
$M_{X(3872)}-M_{D^0}-M_{D^{*0} }=(-0.12\pm0.24)$~MeV~\cite{TheBABAR:2013dja},
which leads  to speculations of the $X(3872)$  as a hadronic
molecule---either a $D\bar D^*$ loosely bound
state~\cite{Tornqvist:2004qy}
or a virtual state~\cite{Hanhart:2007yq}.
Furthermore, a large isospin breaking is found in  its decays:
the process $X(3872) \to J/\psi \pi^+\pi^-$ via a virtual $\rho^0$   and the
process $X(3872)\to J/\psi \pi^+\pi^-\pi^0$ via a virtual $\omega$  have
similar partial widths~\cite{Beringer:1900zz}.  Evidence for  different rates
of charged and neutral $B$ decays  into $X(3872)$ was also found~\cite{Aubert:2008gu}.

These facts have stimulated great interest in  understanding the nature,
production and decays of  the $X(3872)$. An important aspect involves the
discrimination of a compact multiquark configuration and a loosely bound
hadronic molecule configuration. Recent calculations of the
hadroproduction rates at the LHC based on nonrelativistic QCD indicate that
the $X(3872)$ could hardly be an ordinary charmonium
$\chi_{c1}(2P)$~\cite{Butenschoen:2013pxa,Meng:2013gga}, while there are sizable disagreements
in theoretical predictions in the molecule
picture~\cite{Suzuki:2005ha,Bignamini:2009sk,Artoisenet:2009wk,
Artoisenet:2010uu,Esposito:2013ada}.

To clarify the intriguing properties and finally decipher the internal nature,
more accurate data and new processes involving the production and decays of the $X(3872)$
will be helpful. For instance, one may obtain useful information on the flavor
content of the $X(3872)$ from precise measurements of decays of
neutral/charged $B$ mesons into the
$X(3872)$ associated with neutral/charged $K^*$ mesons.

On the other hand,  it is also expedient  to look for the possible analogue of
the $X(3872)$ in the bottom
sector, referred to as $X_b$  following the notation suggested in
Ref.~\cite{Hou:2006it}. If such a state exists,   measurements of its properties
would assist  us  in  understanding the formation of the $X(3872)$ as the
underlying interaction is expected to respect heavy flavor symmetry. In fact,
the existence of such a state was predicted in both the tetraquark
model~\cite{Ali:2009pi} and hadronic molecular
calculations~\cite{Tornqvist:1993ng,Guo:2013sya,Karliner:2013dqa}. The mass of
the lowest-lying $1^{++}$ $\bar b \bar q bq$ tetraquark was predicted to be
10504~MeV in Ref.~\cite{Ali:2009pi}, while the mass of the $B\bar B^*$
molecule based on the mass of the $X(3872)$ is a few tens of MeV
higher~\cite{Guo:2013sya,Karliner:2013dqa}. In Ref.~\cite{Guo:2013sya}, the
mass was predicted to be $(10580^{+9}_{-8})$~MeV for a typical cut-off,
corresponding to a binding
energy of $(24^{+8}_{-9})$~MeV.

Notice that there is a big difference between the predicted $X_b$ and
the $X(3872)$. The distance of the mass of the $X(3872)$ to the $D^0\bar D^{*0}$
threshold is much smaller than the distance to the $D^+D^{*-}$ threshold. This
difference leaves its imprint in the wave function at short distances through
the charmed meson loops so that a sizeble isospin breaking effect
is expected. However, the mass difference between the
charged and neutral $B$ mesons is only
$(0.32\pm0.06)$~MeV~\cite{Beringer:1900zz}, and the binding energy of the $B\bar
B^*$ system may be larger than that in the charmed sector due to a larger
reduced mass. In addition, while the isospin breaking observed in the $X(3872)$
decays into $J/\psi$ and two/three pions can be largely explained by the phase
space difference between the $X(3872)\to
J/\psi\rho$ and the $X(3872)\to J/\psi\omega$~\cite{Gamermann:2009fv}, the
phase space difference between the
$\Upsilon\rho$ and $\Upsilon\omega$ systems will be negligible since the mass
splitting between the $X_b$ and the $\Upsilon(1S)$ is definitely larger than
1~GeV. Therefore, we expect that the isospin breaking effects would be much
smaller for the $X_b$ than that for the $X(3872)$. Consequently, the $X_b$
should be an isosinglet state to a very good approximation, in line with the
predictions in Refs.~\cite{Tornqvist:1993ng,Guo:2013sya,Karliner:2013dqa}.

Since the mass of the $X_b$ is larger than 10~GeV and its quantum numbers
$J^{PC}$ are $1^{++}$, it is unlikely to be discovered at the current
electron-positron colliders, though the prospect for an observation  in the
$\Upsilon(5S,6S)$ radiative decays at the Super KEKB in future may be bright due
to the expected large data sets, of order $50~{\rm
ab}^{-1}$~\cite{Aushev:2010bq}. See Ref.~\cite{He:2014sqj} for a recent search in the $\Upsilon\omega$ final state.  There have been works on the production of the
exotic states, especailly hadronic molecules, at hadron
colliders~\cite{Bignamini:2009sk,Artoisenet:2009wk,
Artoisenet:2010uu,Esposito:2013ada,Ali:2011qi,Ali:2013xba,Guo:2013ufa,Guo:2014ppa}.
In this paper, we will follow closely Ref.~\cite{Guo:2014ppa}, which uses
effective field theory (EFT) to cope with the two-body hadronic final state
interaction (FSI), and focus primarily on the production of the $X_b$ and its
spin partner, a $B^*\bar B^*$ molecule with $J^{PC}= 2^{++}$, denoted as
$X_{b2}$,  at the  LHC and the Tevatron. Results on the production of the spin
partner of the $X(3872)$, $X_{c2}$ with $J^{PC}=2^{++}$, will also be given.
Notice that due to heavy quark spin symmetry, the binding energies of the
$X_{b2}$ and $X_{c2}$ are similar to those of the $X_b$ and $X(3872)$,
respectively. In addition, we will also revisit the production of the $X(3872)$,
and compare the obtained results with the experimental data.

This paper is organized as follows. We begin in  Sec.~\ref{sec:hadroproduction}
by discussing the factorization formula for the $pp/\bar p\to X$ (here $X$ is
used to represent all the above mentioned candidates of hadronic molecules, and
both $pp$ and $p\bar p$ will be written as $pp$ for simplicity in the following)
amplitudes in case that the $X$ states are  bound states not far from the
corresponding thresholds.  Our numerical results for the cross sections are
presented in Sec.~\ref{sec:results}. The last section contains a brief summary.

\section{Hadroproduction}
\label{sec:hadroproduction}

The universal scattering amplitude of particles with short-range interaction
provides an easy way to derive the formula for estimating the cross section of
the inclusive production of an $S$-wave loosely bound hadronic molecule~\cite{
Artoisenet:2009wk,Artoisenet:2010uu}. However, the amplitude derived in an EFT
can also be used for such a purpose~\cite{Guo:2014ppa}. Furthermore, by
investigating the consequences of heavy quark symmetries on the $X(3872)$ within
an EFT framework, Ref.~\cite{Guo:2013sya} predicted the bottom analogues and
the spin partner of $X(3872)$. In the following, we will follow
Ref.~\cite{Guo:2014ppa} and use the EFT as used in Ref.~\cite{Guo:2013sya} to
obtain a factorization formula, which will enable us to estimate the inclusive
production cross sections for the $X$ production.

\begin{figure}[b]
\centering
\includegraphics[width=0.42\linewidth]{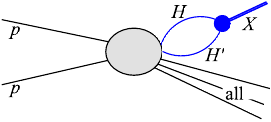}
\caption{ The mechanism considered in the paper for the inclusive production of
the $X$ as a $HH^{\prime}$ bound state in proton--proton collisions. Here, $all$
denotes all the produced particles other than the $H$ and $H^{\prime}$ in the
collision. }
\label{fig:prod}
\end{figure}

When the binding energy of a bound state is small, we can assume that the
formation of the hadronic molecule, which is a long-distance process, would
occur after the production of its constituents, which is of short-distance
nature.
The mechanism is shown in Fig.~\ref{fig:prod}. Therefore, the amplitude for the
production of the hadronic molecule can be written as~\cite{Guo:2014ppa}
\begin{eqnarray}
{\cal M}[X]={\cal M}[HH^{\prime} +\text{all}] \times G \times T_X,
\label{eq:totalamplitude}
\end{eqnarray}
where ${\cal M}[HH^{\prime} + \text{all}]$ is the amplitude for the inclusive
production of heavy mesons $H$ and $H^{\prime}$, $T_X$ is amplitude for the
process $HH'\to X$, and $G$ is the Green function of the heavy meson pair.
In general, the above equation is an integral equation with
all the parts on the right-hand-side involved in an integral over the momentum
of the intermediate mesons. However, in the case that the hadronic molecule is a
loosely bound state, $T_X$ can be approximated by the coupling constant $g$ of
the $X$ to its constituents, and as argued in Ref.~\cite{Artoisenet:2009wk},
one should be able to approximate the production amplitude
${\cal M}[HH^{\prime} + \text{all}]$, which does not take into account the FSI carrying
a strong momentum dependence near threshold, by a constant.
Thus, both ${\cal M}[HH^{\prime} + \text{all}]$ and $g$ can be taken outside the
momentum integral, and $G$ becomes a two-point scalar loop function.

The general differential Monte Carlo (MC) cross section formula for the
inclusive $HH^{\prime}$ production reads
\begin{eqnarray}
	d\sigma[HH^{\prime}(k)]_\text{MC} &=&
	 K_{HH^{\prime}}\frac{1}{\rm flux} \sum_\text{all} \int d \phi_{HH^{\prime} +
	 \text{all}} |{\cal M}[HH^{\prime}(k) +\text{all}]|^2 \frac{d^3k}{(2 \pi)^3 2
	 \mu}.
	 \label{eq:generalProduction}
\end{eqnarray}
where $k$ is the three-momentum in the center-of-mass frame of the $HH^{\prime}$
pair, $\mu$ is the reduced mass of the $HH^{\prime}$ pair and $K_{HH^{\prime}}\sim{\cal O}(1)$ is
introduced because of the overall difference between MC simulation and the
experimental data, while for an order-of-magnitude estimate we can roughly take
$K_{HH^{\prime}}\simeq1$. Without considering the FSI, the matrix element ${\cal M}[HH^{\prime}(k) +\text{all}]$ is a  constant and thus we have: 
\begin{eqnarray}
\frac{d\sigma[HH^{\prime}(k)]_\text{MC}}{dk} \approx k^2.  \label{eq:leading_order_behaviour}
\end{eqnarray}  
On the other hand, the cross section for the
production of the $X$, which stands for $X(3872)$, $X_b$, $X_{b2}$ or $X_{c2}$,
is
\begin{eqnarray}
	\sigma[X] &=&
	 \frac{1}{\rm flux} \sum_\text{all} \int d \phi_{X + \text{all}}
	 \left|{\cal M}[X + \text{all}] \right|^2,
	 \label{eq:generalProductionX}
\end{eqnarray}
where the phase space integration is the same as that in
Eq.~\eqref{eq:generalProduction}.
Therefore the cross section of $X$ can be rewritten with
Eqs.~\eqref{eq:totalamplitude} and~\eqref{eq:generalProduction} as
\begin{eqnarray}
	\sigma[X] &=&   \frac{1}{4m_Hm_{H'}}
	 g^2 |G|^2
\bigg(\frac{d\sigma[HH^{\prime}(k)]}{dk}\bigg)_\text{MC}\frac{4\pi^{2}\mu}{k^{2}}.
\label{eq:cross-section-final}
\end{eqnarray}
Since we will study the production of the hadronic molecules predicted in
Ref.~\cite{Guo:2013sya}, we will use the same Gaussian cutoff to regularize the
divergent loop integral $G$, and have~\cite{Nieves:2012tt}
\begin{eqnarray}
	G(E,\Lambda) &=&  -\frac{\mu}{\pi^2}\bigg[\sqrt{2\pi}\,\frac{\Lambda}{4}+\sqrt{\pi}\,\gamma
	D\left(\frac{\sqrt{2}\gamma}{\Lambda}\right)-\frac{\pi}{2}\,\gamma\,e^{2\gamma^2/\Lambda^2}\bigg],
\label{eq:loopintegral}
\end{eqnarray}
where $D(x)=e^{x^2}\,\int^{x}_0\,e^{-y^2}\,dy$ is the Dawson function,
$\gamma=\sqrt{-2\mu(E-m_H-m_{H'})}$ is the binding momentum and $\Lambda$ is the
cutoff. Following Ref.~\cite{Guo:2013sya}, a range of $[0.5,1.0]$~GeV
will be used to the cutoff $\Lambda$.
By considering only the leading order
contribution, the pole of the bound state satisfies the equation
$1-C_0\,G[E_\text{pole},\Lambda]=0$, where $C_0$ is the leading order low energy
constant which describes the contact interaction between the considered heavy
meson pair.
The renormalization group invariance requires that $C_0$ depends on $\Lambda$ as
well in order to make the physical observables cutoff independent. The coupling
constant $g$ in Eq.~\eqref{eq:cross-section-final} is related to the residue of
the bound state pole by
\begin{eqnarray}
    g^2 = \lim_{s\to s_\text{pole} } (s-M_X^2 )\frac{C_{0}(\Lambda)}
           {1 - C_{0}(\Lambda)\,G(\sqrt{s},\Lambda)}
        =\frac{C_{0}(\Lambda)}
           {d[1 -
           C_{0}(\Lambda)\,G(\sqrt{s},\Lambda)]/ds}\bigg|_{s=M_X^2},
\label{eq:coupling}
\end{eqnarray}
where $s$ is the center-of-mass energy squared.

\section{Results and discussions}
\label{sec:results}

In order to form a
molecule, the mesonic constituents must be produced at first and  have to move
collinearly with a small relative momentum. Such configurations  originate
from the inclusive  QCD  process which contains a  $\bar QQ$
pair with a similar relative momentum in the final state. Thus, at least
a third parton needs to be produced in the recoil direction, which
corresponds to a $2\to3$ parton process. In our explicit realization,  the
$2\to 3$ process can be generated initially  through  hard scattering,  and
the parton shower will produce more quarks via soft radiations.

Following our previous work~\cite{Guo:2013ufa}, we use
Madgraph~\cite{Alwall:2011uj} to generate the  $2\to 3$  partonic events with a
pair of a heavy quark and an antiquark ($\bar b b$ or $\bar cc$) in the final
states, and then pass them to the MC event generators for
hadronization.
We choose Herwig~\cite{Bahr:2008pv} and
Pythia~\cite{Sjostrand:2007gs} as the hadronization generators, whose outputs
are analyzed using the Rivet library~\cite{Buckley:2010ar}.

To improve the efficiency of the calculation, we
apply the partonic cuts for the transverse momentum $p_T>2$~GeV for heavy quarks
and light jets, $m_{c\bar c}< 4.5$~GeV  ($k_{D\bar D^*}=1.14$~GeV and $k_{D^*\bar D^*}=1.02$~GeV),   $m_{b\bar b}< 10.7$~GeV ($k_{B\bar B^*}=715$~MeV and $k_{B^*\bar B^*}=517$~MeV at the hadron level),  and $\Delta R(c, \bar c)<1$($\Delta R(b, \bar b)<1$)  where  $\Delta R=\sqrt{\Delta \eta^2 +\Delta \phi^2}$ ($\Delta \phi$ is
the azimuthal angle difference and $\Delta \eta$ is the pseudo-rapidity
difference of the $b\bar b$).

Before proceeding to the predictions for the bottom anologues and
the spin partner of the $X(3872)$, we shall revisit the production of the
$X(3872)$, and compare the results with the experimental data.
Such a  comparison requires a range for the branching ratio
${\cal B}(X(3872)\to J/\psi \pi^+ \pi^-)$.  Making use of the Babar upper limit
for ${\cal B}(B^+\to X(3872)K^+)$~\cite{Aubert:2005vi} and the most recent Belle
measurement of ${\cal B}(B^+\to X(3872)K^+) \times {\cal B}(X(3872)\to J/\psi
\pi^+ \pi^-)$~\cite{Choi:2011fc},
\begin{eqnarray}
 {\cal B}(B^+\to X(3872)K^+) &<& 3.2\times 10^{-4}, \nonumber\\
{\cal B}(B^+\to X(3872)K^+) \times {\cal B}(X(3872)\to J/\psi \pi^+ \pi^-) &=& (8.63\pm0.82\pm0.52)\times 10^{-4},
\end{eqnarray}
we can derive a lower bound:
\begin{eqnarray}
{\cal B}(X(3872)\to J/\psi \pi^+ \pi^-)> 0.027.
\end{eqnarray}
On the other hand, summing over the branching fractions of $X(3872)$ to all
measured channels which, in addition to the
$J/\psi\pi^+\pi^-$~\cite{Choi:2011fc}, include $D^0\bar D^{*0}+c.c.$~\cite{Adachi:2008sua},
$J/\psi\omega$~\cite{delAmoSanchez:2010jr}, $\psi'\gamma$ and
$J/\psi\gamma$~\cite{Aubert:2008ae,Bhardwaj:2011dj} can
provide an upper bound for the  branching fraction of the $X(3872)\to J/\psi
\pi^+ \pi^-$:
\begin{eqnarray}
{\cal B}(X(3872)\to J/\psi \pi^+ \pi^-)<0.083
\end{eqnarray}

In Tab.~\ref{tab:X_3872_compare}, we show the integrated cross sections
(in units of nb) for the  $pp/\bar p\to X(3872)$ and
compare with previous theoretical
estimates~\cite{Bignamini:2009sk,Artoisenet:2009wk} and  experimental
measurements by the CDF Collaboration~\cite{Bauer:2004bc}
\begin{eqnarray}
{ \sigma}(p\bar p\to X) \times {\cal B}(X(3872)\to
J/\psi\pi^+\pi^-)=(3.1\pm0.7)~{\rm nb},
\end{eqnarray}
and  by the CMS Collaboration~\cite{Chatrchyan:2013cld}
\begin{eqnarray}
{ \sigma}(pp\to X) \times {\cal B}(X(3872)\to
J/\psi\pi^+\pi^-)=(1.06\pm0.11\pm0.15)~{\rm nb}.
\end{eqnarray}
The same kinematical cuts on the transverse momentum and
rapidity as those in the experimental analyses were implemented: $p_T>5 $ GeV
and $|y|<1.2$ at the Tevatron and $10 {\rm GeV}<p_T<50 {\rm GeV}$ and $|y|<0.6$
at  the LHC with $\sqrt{s}=7$ TeV.
In this table,  we have  converted the experimental data to ${ \sigma}(p\bar
p/pp\to X)$.
A very small upper bound was derived for ${ \sigma}(p\bar
p/pp\to X)$ in Ref.~\cite{Bignamini:2009sk}, and the
predicted values are increased in Ref.~\cite{Artoisenet:2009wk} by taking into
account the FSI using the universal scattering amplitude. As shown in this
table, our results agree with the experimental measurements quite well,
which validates our calculation based on an EFT treatment of the FSI.

Uncertainties in our results  come from the parameter $\Lambda$ in the loop
function  in Eq.~\eqref{eq:cross-section-final}. Based on heavy quark
symmetries, this parameter has been adopted as
$\Lambda\in[0.5,1]$~GeV~\cite{Guo:2013sya}.  Different values will give rise to different binding energies of  the counterparts for instance  the $X_b$, ranging from $24$ MeV to $66$ MeV. Measurements of the $X_b$ mass    in future will reduce the errors. Taking into account these uncertainties, our results for the cross section at the Tevatron  are given as
\begin{eqnarray}
{ \sigma}(p\bar p\to X(3872)) &=& \left\{\begin{array}{c}
(10, 47)~{\rm nb} \;\;\; {\rm for~Herwig}\\
 \;(7,33)~{\rm nb} \;\;\; {\rm for~Pythia}
\end{array}\right. ,
\end{eqnarray}
and at the LHC with $\sqrt{s}=7$ TeV
\begin{eqnarray}
{ \sigma}(p p\to X(3872)) &=& \left\{\begin{array}{c}
(16, 72)~{\rm nb} \;\;\; {\rm for~Herwig}\\
 \;(7,32)~{\rm nb} \;\;\; {\rm for~Pythia}
\end{array}\right. .
\end{eqnarray}

\begin{table}[t]
\caption{Integrated cross sections (in units of nb) for   $pp/\bar p\to X(3872)$
compared with previous theoretical estimates~\cite{Bignamini:2009sk,
Artoisenet:2009wk} and  experimental measurements by CDF~\cite{Bauer:2004bc}
and CMS~\cite{Chatrchyan:2013cld}.
Results outside (inside) brackets are obtained using Herwig (Pythia).
Kinematical cuts used are: $p_T>5 $ GeV and $|y|<1.2$ at Tevatron and $10~{\rm
GeV}<p_T<50 {\rm GeV}$ and $|y|<0.6$ at LHC with $\sqrt{s}=7$ TeV.  We have
converted the experimental data  ${ \sigma}(p\bar p\to X) \times {\cal
B}(X(3872)\to J/\psi\pi^+\pi^-)=(3.1\pm0.7){\rm nb}$~\cite{Bauer:2004bc} and
${ \sigma}(pp\to X) \times {\cal B}(X(3872)\to J/\psi\pi^+\pi^-)=
(1.06\pm0.11\pm0.15){\rm nb}$~\cite{Chatrchyan:2013cld} into cross sections
using ${\cal B} (X(3872)\to J/\psi \pi^+ \pi^-)\in[0.027,0.083]$ as discussed
in the text.
}
\label{tab:X_3872_compare}
\begin{tabular}{lccccc}
 \hline\hline
 $\sigma(pp/p\bar p\to X(3872))$ &Ref.~\cite{Bignamini:2009sk} & Ref.~\cite{Artoisenet:2009wk} & {  $\Lambda=0.5$ GeV} & { $\Lambda=1$ GeV}  & Experiment  \\\hline
Tevatron& $<0.085$& $1.5$--$23$  &10(7) &    47(33) &
37--115~\cite{Bauer:2004bc}\\
LHC7   &-- & 45--100~\footnote{Estimate based on non-relativistic QCD.}
&16(7) & 72(32) &
13--39~\cite{Chatrchyan:2013cld}\\
\hline\hline
\end{tabular} \end{table}

\begin{figure}[b]
\includegraphics[width=0.49\textwidth]{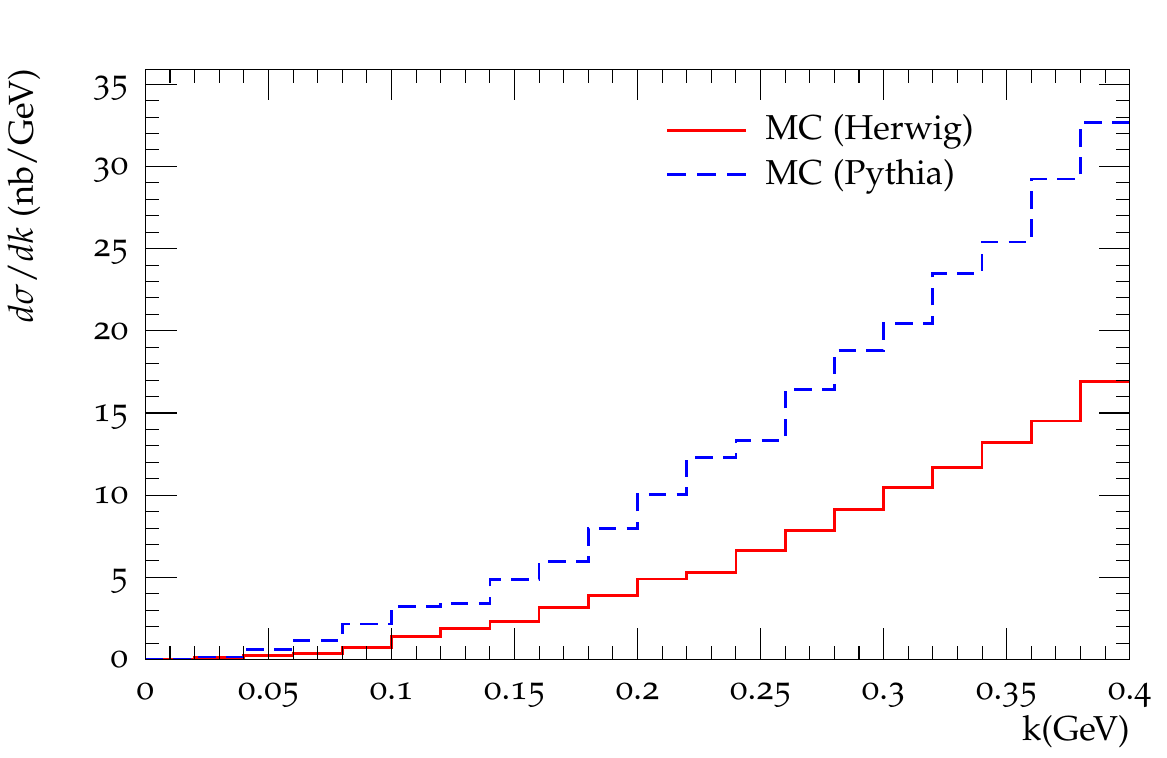}\hfill
\includegraphics[width=0.49\textwidth]{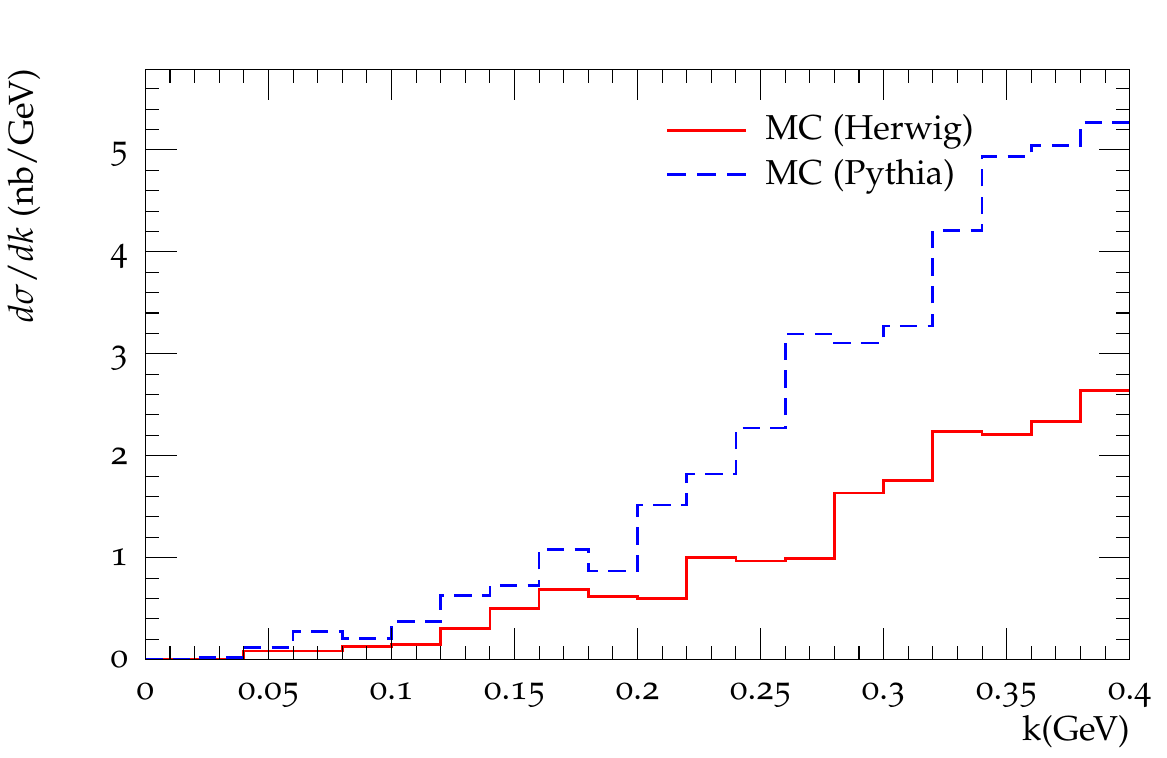}
\includegraphics[width=0.49\textwidth]{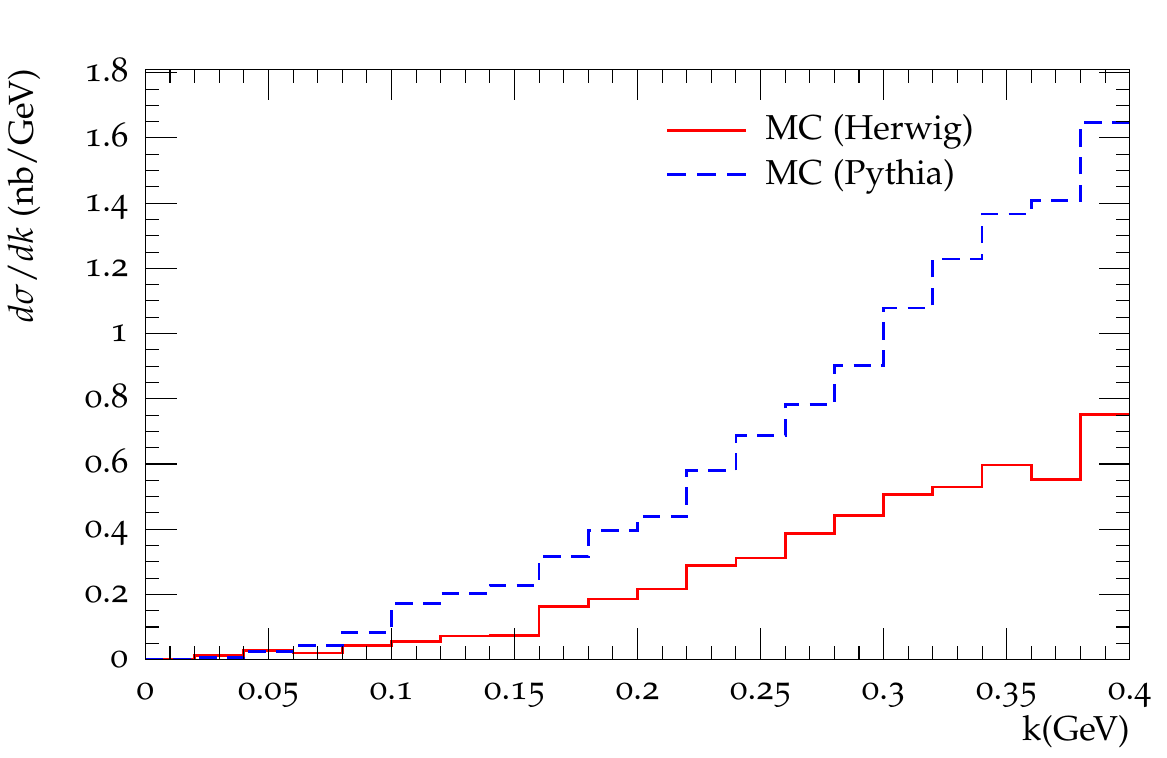}
\caption{\label{fig:mass_LHC}
Differential cross sections ${d\sigma}/{dk}$  (in units of nb/GeV) for  the
process $pp\to B^0\bar B^{*0}$ at the LHC with $\sqrt s=8$ TeV (upper panels) and at the Tevatron
with $\sqrt s=1.96$~TeV (lower panel).  The kinematic cuts for the left-upper panel are  used as
$|y|<2.5$ and $p_T>5$ GeV, which lie in the phase-space regions of the ATLAS
and CMS detectors, for the Tevatron experiments
(CDF and D0) at 1.96~TeV (the lower panel), we use $|y| <0.6$;
the rapidity range $2.0<y<4.5$ is used for  LHCb (the right-upper panel). }
\end{figure}

\begin{figure}[b]
\includegraphics[width=0.49\textwidth]{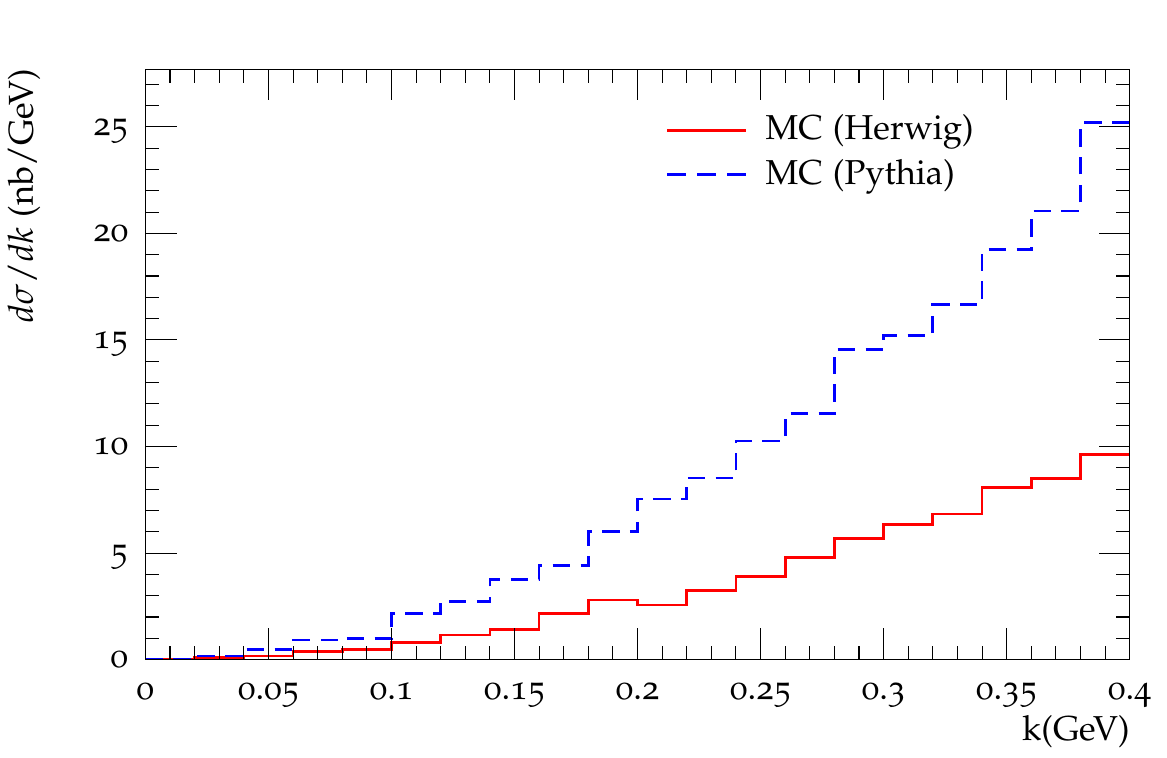}\hfill
\includegraphics[width=0.49\textwidth]{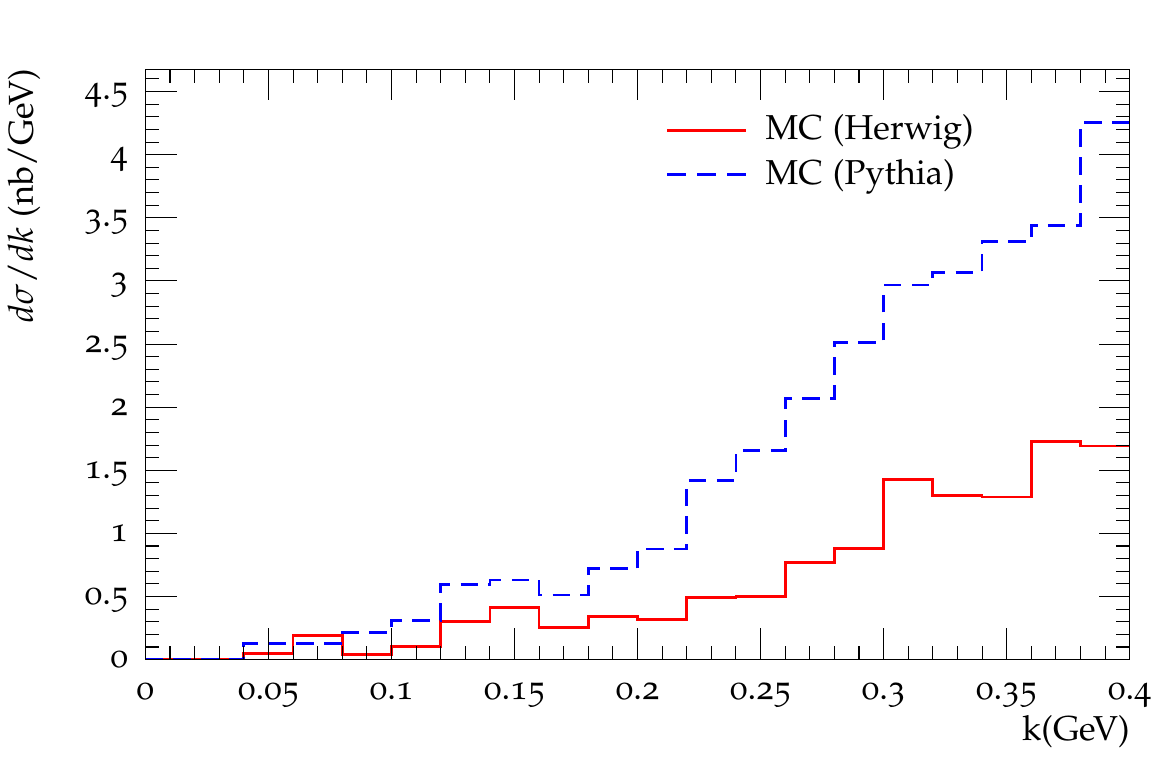}
\includegraphics[width=0.49\textwidth]{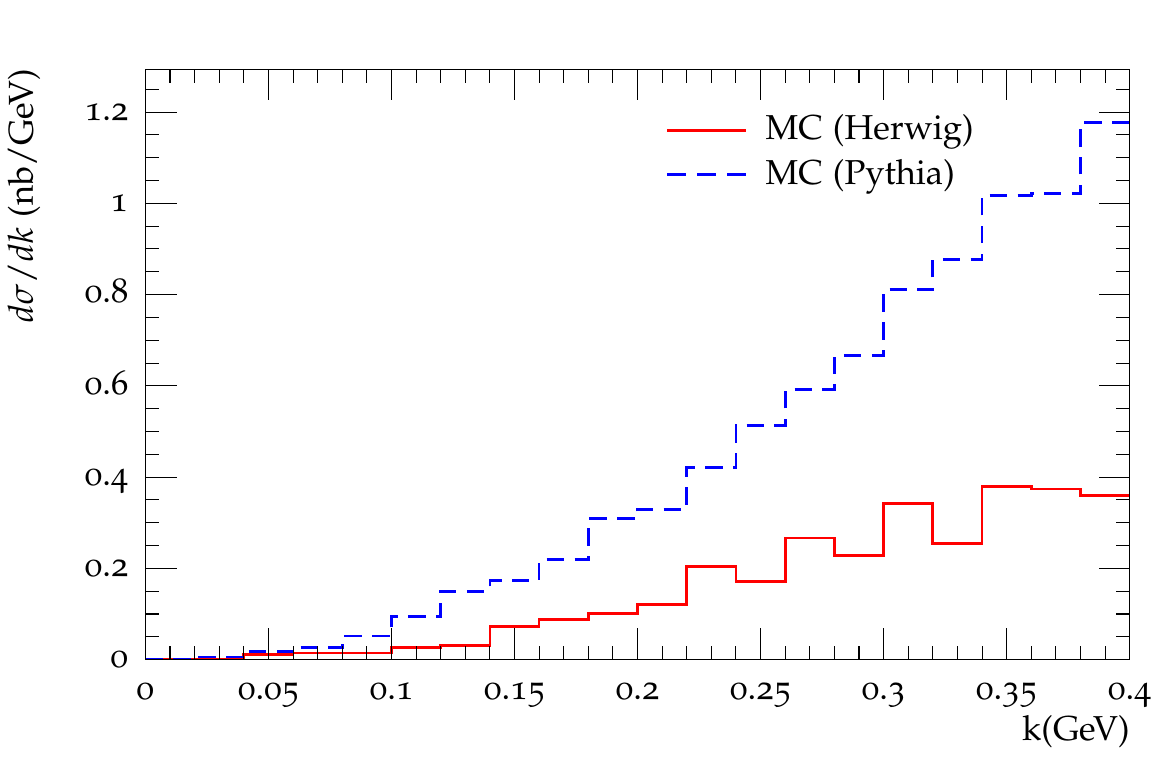}
\caption{\label{fig:mass_LHC_BstarBstar}
Same as Fig.~\ref{fig:mass_LHC}  but for  the $B^*\bar B^*$ final state.}
\end{figure}

Based on $10^7$ partonic events generated by Madgraph, we show the differential
cross sections ${d\sigma}/{dk}$  (in units of nb/GeV) for the process $pp\to B^0\bar
B^{*0}$  in Fig.~\ref{fig:mass_LHC}, and the ones for the reaction $pp\to B^{*0}\bar
B^{*0}$  in Fig.~\ref{fig:mass_LHC_BstarBstar} at the LHC with the
center-of-mass energy $\sqrt s=8$~TeV and at the
Tevatron with $\sqrt s=1.96$~TeV.  The
kinematic cuts  are  $|y|<2.5$ and $p_T>5$~GeV, where $y$ and $p_T$
are the rapidity and the transverse momentum of the bottom mesons, respectively,
which lie in the phase space regions of the ATLAS and CMS detectors. For the
Tevatron experiments (CDF and D0) at 1.96~TeV, we use $|y| <0.6$; the rapidity
range $2.0<y<4.5$ is used for the LHCb detector.  We have checked that $d\sigma/dk$ is
approximately proportional to $k^{2}$, cf. Eq.~\eqref{eq:leading_order_behaviour}. 

\begin{table}[t]
\caption{Integrated cross sections (in units of nb)
for the $pp/\bar p\to X_b$, 
and $pp/\bar p\to X_{b2}$ at the LHC and Tevatron.
Results out of (in) brackets are obtained using Herwig(Pythia).
The rapidity range $|y| <2.5$
has been assumed for the LHC experiments 
(ATLAS and CMS) at 7, 8 and 14 TeV; 
for the Tevatron experiments 
(CDF and D0) at 1.96 TeV, we use $|y| <0.6$;
the rapidity range $2.0<y<4.5$ is used for the LHCb. }
\label{tab:integratedCrossSection}
\begin{tabular}{cccccc}
 \hline\hline
 $X_b$ & $E_{X_b}=24$ MeV($\Lambda=0.5$ GeV) & $E_{X_b}=66$ MeV($\Lambda=1$\
 
  GeV) \\\hline
Tevatron
 &0.08(0.18)
 &0.61(1.4)
\\
LHC 7
 &1.5(3.1)
 &12(23)
\\
LHCb 7
 &0.25(0.49)
 &1.9(3.7)
\\
LHC 8
 &1.8(3.6)
 &14(27)
\\
LHCb 8
 &0.3(0.62)
 &2.2(4.7)
\\
LHC 14
 &3.2(6.8)
 &24(51)
\\
LHCb 14
 &0.65(1.3)
 &4.9(9.7)
\\
 \hline
 $X_{b2}$ & $E_{X_{b2}}=24$ MeV($\Lambda=0.5$ GeV) & $E_{X_{b2}}=66$\
 
  MeV($\Lambda=1$ GeV) \\\hline
Tevatron
 &0.05(0.13)
 &0.36(1.)
\\
LHC 7
 &0.92(2.3)
 &6.9(17)
\\
LHCb 7
 &0.14(0.36)
 &1.1(2.7)
\\
LHC 8
 &1.1(2.7)
 &8.1(20)
\\
LHCb 8
 &0.19(0.46)
 &1.4(3.5)
\\
LHC 14
 &1.9(5.)
 &15(37)
\\
LHCb 14
 &0.38(0.96)
 &2.9(7.2)
\\
 \hline\hline
%
 $X_{c2}$ & $E_{X_{c2}}=4.8$ MeV($\Lambda=0.5$ GeV) & $E_{X_{c2}}=5.6$\
 
  MeV($\Lambda=1$ GeV) \\\hline
Tevatron
 &4.4(3.)
 &22(15)
\\
LHC 7
 &66(44)
 &327(216)
\\
LHCb 7
 &14(8.5)
 &71(42)
\\
LHC 8
 &74(52)
 &369(256)
\\
LHCb 8
 &17(10)
 &83(50)
\\
LHC 14
 &135(90)
 &672(446)
\\
LHCb 14
 &35(19)
 &174(92)
\\
 \hline\hline
\end{tabular} \end{table}



Integrated cross
sections (in units of nb) for the $pp\to X_b$, and $pp\to X_{b2,c2}$ are collected
in Tab.~\ref{tab:integratedCrossSection}. Results outside (inside) brackets are
obtained using Herwig (Pythia).
 From the table, one sees that the
cross sections for the $X_{b2}$ is similar to those for the $X_b$, and the ones
for the $X_{c2}$ are of the same order as those for the $X(3872)$ given in
Table~\ref{tab:X_3872_compare} and are two orders of magnitude larger than those
for their bottom analogues.

Recently, the CMS Collaboration has presented  results of a first search
for  new bottomonium states, with the main focus on the $X_b$, decaying to
$\Upsilon(1S)\pi^+\pi^-$. The search is based on a data sample corresponding to
an integrated luminosity of 20.7~${\rm fb}^{-1}$ at
$\sqrt{s} = 8\,$TeV ~\cite{CMSconstraint}.  No evidence for the  $X_b$ is
found, and the upper
limit at a confidence level of 95\% on the product of the production cross
section of the $X_b$ and the decay branching fraction of $X_b\to
\Upsilon(1S)\pi^+\pi^-$ has been set to be
\begin{eqnarray}
 \frac{\sigma(pp\to X_b\to \Upsilon(1S)\pi^+\pi^-)}{\sigma(pp\to
\Upsilon(2S)\to
\Upsilon(1S)\pi^+\pi^-)} < (0.009, 0.054)~, \label{eq:CMSratio}
\end{eqnarray}
where the range corresponds to the variation of the $X_b$ mass from 10 to
11~GeV.

Using the current experimental data on the $\sigma(pp\to \Upsilon(2S))$, we can
convert the above ratio into the cross section which can be directly
compared with our results. Since the masses of the $\Upsilon(2S)$ and $X_b$ are
not very different,  it may be a good approximation to assume that the   ratio
given in Eq.~\eqref{eq:CMSratio} is insensitive to kinematic cuts. Using the
CMS measurement in Ref.~\cite{Chatrchyan:2013yna}:
\begin{eqnarray}
 \sigma (pp\to \Upsilon(2S) ) {\cal B}(\Upsilon(2S)\to \mu^+\mu^-) 
  = (2.21\pm 0.03^{+0.16}_{-0.14}\pm 0.09)~{\rm nb},
\end{eqnarray}
with  the cuts $p_T< $50 GeV and $|y|<2.4$ for the $\Upsilon(2S)$,
we get
\begin{eqnarray}
 \sigma (pp\to X_b ) {\cal B}(X_b\to\Upsilon(1S) \pi^+\pi^-) 
  < (0.18, 1.11)~{\rm nb}. \label{eq:cmsUpperBound}
\end{eqnarray}
Taking into account   theoretical  errors, our estimate for the cross section $\sigma
(pp\to X_b )$ is
\begin{eqnarray}
 \sigma (pp\to X_b )\sim \left\{\begin{array} {c}
 (1.8, 14)~{\rm nb}\;\;\;{\rm for~Herwig}\\
 (3.6, 27)~{\rm nb}\;\;\;{\rm for~Pythia}
 \end{array}\right. . \label{eq:pptoXbResults}
\end{eqnarray}
However, since the branching ratio ${\cal B}(X_b\to \Upsilon(1S)\pi^+\pi^-)$ is
expected to be tiny because of isospin breaking (see below),  our
result given in Eq.~\eqref{eq:pptoXbResults} is consistent with the CMS upper bound in Eq.~\eqref{eq:cmsUpperBound}.

As already discussed in the Introduction, the $X_b$ and $X_{b2}$ are
isosinglets. In contrast to the $X(3872)$, the isospin breaking decays of
these two states will be heavily suppressed. Thus, one shall not simply make an
analogy to the $X(3872)\to J/\psi\pi^+\pi^-$ and attempt to search for the $X_b$ in
the $\Upsilon(1S,2S,3S)\pi^+\pi^-$ channels, as the isospin of the
$\Upsilon(1S,2S,3S)\pi^+\pi^-$ systems is one when the quantum numbers
are $J^{PC}=1^{++}$. This could be the reason for the negative search result
by the CMS Collaboration~\cite{CMSconstraint}. Possible channels which can be used
to search for the $X_b$ and $X_{b2}$ include the
$\Upsilon(nS)\gamma\,(n=1,2,3)$, $\Upsilon(1S)\pi^+\pi^-\pi^0$ and
$\chi_{bJ}\pi^+\pi^-$. The $X_{b2}$ can also decay into $B\bar B$ in a $D$-wave,
and the decays of the $X_{c2}$ are similar to those of the $X_{b2}$ with the
bottom being replaced by its charm analogue. The isospin
breaking decay  $X_{c2}\to J/\psi\pi^+\pi^-$ through an intermediate $\rho$
meson should be largely suppressed compared with the decay of the $X(3872)$ into
the same particles because the mass of the $X_{c2} $ is about 140~MeV higher
than that of the $X(3872) $, and the phase space difference between the
$J/\psi\rho$ and $J/\psi\omega$ becomes negligible.


Compared with the pionic decays, the $\Upsilon(nS)\gamma\,(n=1,2,3)$ final
states are advantageous because no pion needs to be disentangled  from the
combinatorial background. The disadvantage is the low efficiency in
reconstructing a photon at hadron colliders. Since the $X(3872)$ meson has a
sizable partial decay width into the $J/\psi \gamma$~\cite{Beringer:1900zz}
\begin{eqnarray}
 {\cal B}(X(3872)\to \gamma J/\psi) > 6\times 10^{-3},
\end{eqnarray}
presumably the branching ratio for the $X_b\to \gamma \Upsilon $  is of this
order and see Ref.~\cite{Li:2014uia} for an estimate.  If so,  the cross section for the
$pp\to X_b\to \gamma \Upsilon(1S)\to \gamma \mu^+\mu^-$ is of ${\cal
O}(10~\text{fb})$ or even larger when summing up the $\Upsilon(1S,2S,3S)$. Since
the CMS and ATLAS Collaborations have accumulated more than
20~fb$^{-1}$ data~\cite{ATLAS:luminosity,CMS:luminosity}, we expect
at least a few hundred events. Less events will be collected at the LHCb
detector due to a smaller integrated luminosity, ${\cal O}(3~{\rm
fb}^{-1})$~\cite{LHCb:luminosity}. Nevertheless, the future prospect  is
bright since a data sample of about $3000~{\rm fb}^{-1}$, will be collected,
for instance, by ATLAS after the upgrade~\cite{ATLAS:2013hta}.

Apart from the production rates,  the nonresonant background contributions can
also play an important role in  the search for these molecular states at hadron
colliders since a signal could be buried by a huge background.  To
investigate this issue, we consider  the $X_b$ as an example, which
will be reconstructed in $\Upsilon+\gamma$ final states.  In this process,  the
inclusive  cross section $\sigma(pp\to  \Upsilon)$  can
serve as an upper bound for the background. It has been measured  at $\sqrt s= 7 $~TeV
by the ATLAS Collaboration as~\cite{Aad:2012dlq}
\begin{eqnarray}
\sigma(pp\to  \Upsilon(1S) (\to \mu^+\mu^-))= (8.01\pm 0.02\pm
0.36\pm 0.31)~\text{nb},
\end{eqnarray}
with  $p_T< 70 $ GeV and $|y|<2.25$.
Our results in Tab.~\ref{tab:integratedCrossSection} show that the corresponding  cross section
for the $pp\to  X_b$  is about 1~nb at $\sqrt s= 7 $~TeV.  It is noteworthy to
point out that our kinematic cuts in $p_T$ are more stringent compared to the
ones set by the ATLAS Collaboration.   Using the
integrated luminosity in
2012, 22~fb$^{-1}$~\cite{ATLAS:luminosity}, we have  a lower bound  estimate for the
signal/background ratio
\begin{eqnarray}
 \frac{S}{\sqrt{B} } \gtrsim \frac{ 1\times  22\times 10^6  \times 2.6\%
\times 10^{-2}} { \sqrt{ 8 \times 22 \times 10^6 }} \simeq 0.4,
\end{eqnarray}
where $2.6\%$ is the branching fraction of the $\Upsilon(1S)\to
\mu^+\mu^-$~\cite{Beringer:1900zz}, and $10^{-2}$ is a rough estimate for the
branching fraction of the $X_b\to \Upsilon(1S)\gamma$.   The value of the signal/background ratio can be  significantly  enhanced in the data analysis by
employing suitable kinematic cuts which can  greatly  suppress  the
background, and accumulating many more events based on the upcoming $3000~{\rm
fb}^{-1}$ data~~\cite{ATLAS:2013hta}.

\section{Summary}
In summary, we have made use of  the Monte Carlo event generator tools Pythia
and Herwig, and explored the inclusive processes $pp/\bar p\to B^0\bar B^{*0}$
and $pp/\bar p\to B^{*0}\bar B^{*0}$ at  hadron colliders.  Based on the
molecular picture, we have derived an order-of-magnitude  estimate for the production rates of the
$X_b$, $X_{b2}$ and $X_{c2}$ states, the bottom and spin partners of the
$X(3872)$, at the LHC and Tevatron experiments.  We found
that the cross sections are at the nb level for the hidden bottom hadronic
molecules $X_b$ and $X_{b2}$, and two orders of magnitude larger for the
$X_{c2}$. Therefore, one should be able to observe them at hadron colliders if
they exist in the form discussed here.
The channels which can be used to search for the $X_b$ and $X_{b2}$ include the
$\Upsilon(nS)\gamma\,(n=1,2,3)$, $\Upsilon(1S)\pi^+\pi^-\pi^0$,
$\chi_{bJ}\pi^+\pi^-$ and $B\bar B$ (the last one is only for the $X_{b2}$),
and the channels for the $X_{c2}$ is similar to those for the $X_{b2}$ (with
the bottom replaced by its charm analogue). In
fact, both the ATLAS and D0 Collaborations reported an observation of the
$\chi_b(3P)$~\cite{Aad:2011ih,Abazov:2012gh}, whose mass is
$(10534\pm9)$~MeV~\cite{Beringer:1900zz}, slightly lower than the $X_b$ and
$X_{b2}$, in the $\Upsilon(1S,2S)\gamma$ channels. A search for these
states will provide very useful information in understanding the $X(3872)$ and
the interactions between heavy mesons. Especially, if the $X_b$, which is the
most robust among the predictions in Ref.~\cite{Guo:2013sya} based on heavy
quark symmetries, cannot be found in any of these channels, it may imply a
non-molecular nature for the $X(3872)$.

\section*{ Acknowledgments}
FKG would like to thank Institute of Theoretical Physics of Chinese Academy
of Sciences, where part of the work was done, for the hospitality. This work is
supported in part by the DFG and the NSFC through funds provided to the Sino-German CRC 110 ``Symmetries and the Emergence of Structure in QCD'', and by the NSFC (Grant No. 11165005).
We also acknowledge the support of the European Community-Research Infrastructure
Integrating Activity ``Study of Strongly Interacting Matter'' (acronym
HadronPhysics3, Grant Agreement n. 283286) under the Seventh Framework Programme of EU.
Tabulated results  of the distributions
for the $pp/\bar p\to B^{(*)}\bar B^*$ and $pp/\bar p\to
D^{(*)}\bar D^*$ can be found  at:
\url{http://www.itkp.uni-bonn.de/~weiwang/hadronLHC.shtml}.

\end{document}